\documentclass[showpacs,twocolumn,showkeys,amsmath,amssymb,pra]{revtex4-1}

\usepackage{graphicx}   

\newcommand{\f}{a}
\newcommand{\xx}{h}

\newcommand{\ri}{{i}}

\begin{document}

\title{Fractional photon-assisted tunneling in an optical superlattice: large contribution to particle transfer} 

\author{Martin Esmann}
\email{martin.esmann@uni-oldenburg.de}
\author{Niklas Teichmann}
\author{Christoph Weiss}

\affiliation{Institut f\"ur Physik, Carl von Ossietzky Universit\"at,
                D-26111 Oldenburg, Germany
}

                  
\date{\today}
 
\begin{abstract}
Fractional photon-assisted tunneling is investigated  both analytically and numerically for few interacting ultra-cold atoms in the double-wells of an optical superlattice. This can be realized experimentally by adding periodic shaking to an existing experimental setup [Phys. Rev. Lett. 101, 090404 (2008)]. Photon-assisted tunneling is visible in the particle transfer between the wells of the individual double wells. In order to understand the physics of the photon-assisted tunneling, an effective model based on the rotating wave approximation is introduced. The validity of this effective approach is tested for wide parameter ranges  which are accessible to experiments in double-well lattices. The effective model goes well beyond previous perturbation theory approaches and is useful to investigate in particular the fractional photon-assisted tunneling resonances. Analytic results on the level of the experimentally realizable two-particle quantum dynamics show very good agreement with the numerical solution of the time-dependent Schr\"odinger equation. Far from being a small effect, both the one-half-photon and the one-third-photon resonance are shown to have large effects on the particle transfer. 
\end{abstract} 
\maketitle 


\section{\label{sec:introduction}Introduction}
Recent experimental developments for ultra-cold atoms in optical lattices allow to count ever smaller numbers of atoms. Ref.~\cite{CheinetEtAl08} uses an optical superlattice which is a double-well lattice~\cite{SebbyStrableyEtAl07, *YukalovYukalova09} to multiply single double-well potentials with less than six atoms at each double well. It now is even possible to count single atoms~\cite{BakrEtAl09, *ShersonEtAl10}. These developments open new possibilities to investigate few-particle quantum dynamics far from the mean-field effects often encountered, e.g., in large Bose-Einstein condensates (BEC). The focus of the present paper will be the influence of periodic shaking on the quantum dynamics of few particles in the double wells of the optical superlattice of Refs.~\cite{CheinetEtAl08,Folling07}. We will use the fact that the experimental parameters can be chosen such that tunneling between neighboring double wells can be discarded~\cite{Foelling10}. The fact that the experimental setup of Ref.~\cite{CheinetEtAl08} can realize few particles per lattice site in a very controlled way is based on the Mott-insulator~\cite{JakschEtAl98, *GreinerEtAl02} with which the experiment starts~\cite{CheinetEtAl08}.

Periodic shaking is already interesting on the single-particle level~\cite{GrifoniHanggi98} where effects ranging from destruction of tunneling~\cite{GrossmannEtAl91,*KierigEtAl08}  over tunneling-control~\cite{Holthaus92,*LuEtAl10} to population transfer between defects~\cite{Weiss06b} and quantum scattering in driven single- and double-barrier systems~\cite{GarttnerEtAl10} have been investigated. On the many-particle level, recent investigations range from transport of bound pairs in optical lattices~\cite{KudoEtAl09, *WeissBreuer09} over many-body coherent destruction of tunneling~\cite{GongEtAl09} and an ac-driven atomic quantum motor~\cite{PonomarevEtAl09} to frustrated quantum antiferromagnetism~\cite{EckardtEtAl10}. Periodically kicked systems have been investigated, e.g.,  in Refs.~\cite{dArcy01, *StrzysEtAl08, *GhoseEtAl08, *CreffieldEtAl06, *TrailEtAl08} and references therein.

One of the interesting aspects of periodic shaking is photon-assisted tunneling~\nocite{KohlerSols03}\cite{KohlerSols03,TeichmannEtAl09}. Both integer photon-assisted tunneling~\cite{SiasEtAl08} and the related effect of dynamic localization~\cite{EckardtEtAl09b} have been investigated experimentally for weakly interacting BECs in optical lattices. The experimentally observed one- and two-photon resonances~\cite{SiasEtAl08} essentially are single-particle effects that survive interactions.

\begin{figure}
\includegraphics[width=0.8\linewidth]{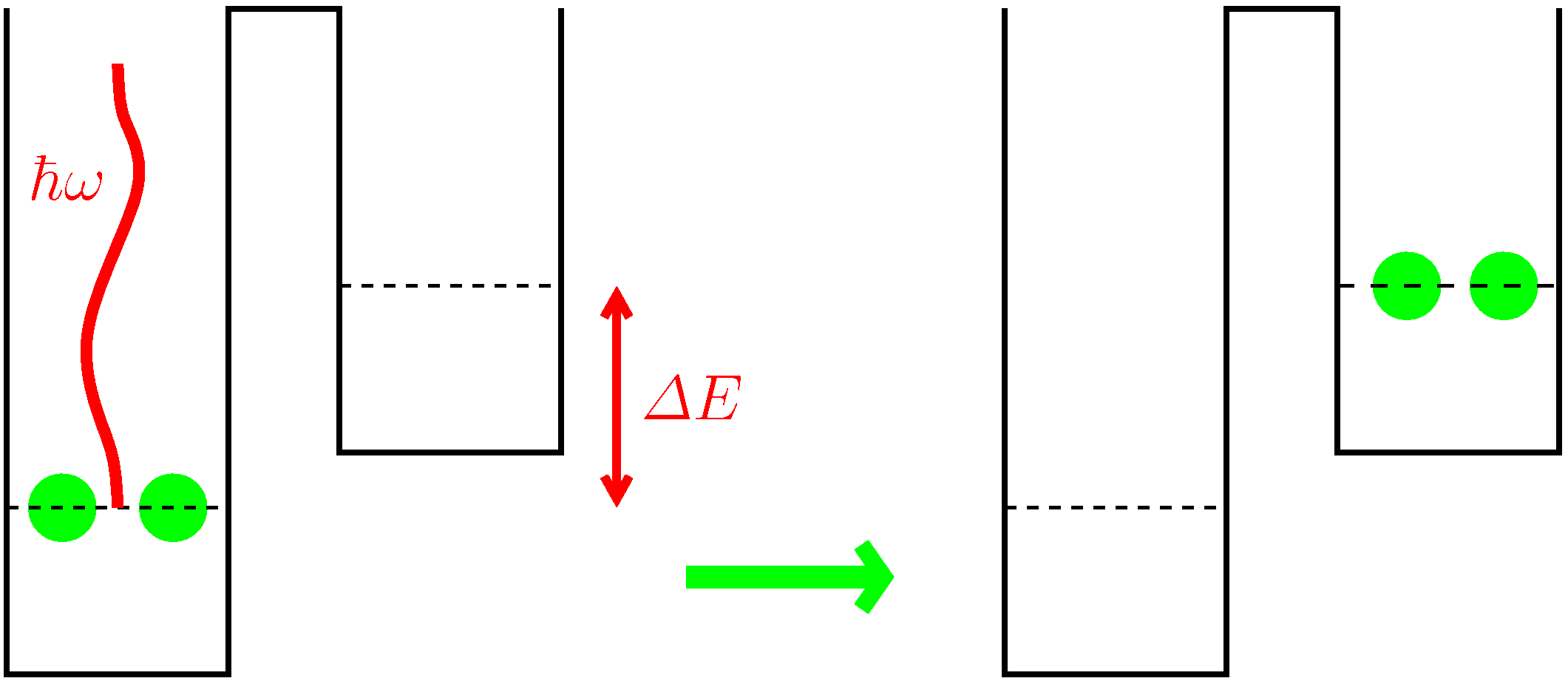}
\caption{\label{fig:schema}(Color online) Schematic drawing of fractional photon-assisted tunneling for the example of a 1/2-photon resonance for the experimentally relevant~\cite{CheinetEtAl08} case of $N=2$ particles. The energy of one ``photon'' is enough to make two particles tunnel. However, the ``photons'' are time-dependent potential modulations in the kilo-hertz-regime rather than real photons which could be absorbed. For ultra-cold atoms in periodically shaken double-well potentials, fractional photon-assisted tunneling is an interaction induced many-particle effect.}
\end{figure}

 Fractional photon-assisted tunneling (cf.\ Fig.~\ref{fig:schema}) is a true interaction driven many-particle effect which has so far been shown to be a small effect both for few particles in double wells~\cite{TeichmannEtAl09,XieEtAl10} and for BECs in double wells~\cite{EckardtEtAl05} as well as for super Bloch oscillations~\cite{HallerEtAl10}.  The aim of the present manuscript is to show that fractional photon-assisted tunneling can, in fact, be a large effect which can be observed with the existing experimental setup~\cite{CheinetEtAl08}. Strictly speaking, there are no photons absorbed in the photon-assisted tunneling in periodically shaken systems: the ``photons'' are modulations of the potential on time-scales in the kilo-hertz regime. There are, however, resonances which correspond to integer or fractional multiples of the energy corresponding to the shaking frequency. The tunneling at those resonances is called photon-assisted tunneling - a half-photon resonance corresponds to one photon transferring two atoms to the other well, a one-third-photon resonance corresponds to one photon transferring three atoms.

The paper is organized as follows: In Sec.~\ref{sec:model}, the model used to describe the interacting particles in a periodically shaken double well is introduced. In order to understand the physics of the time-dependent system, an effective time-independent model is introduced in Sec.~\ref{sec:method}. Section~\ref{sec:results} investigates the quality of the effective approach systematically via both analytic calculations and exact numeric diagonalization of the approximated model. 

\section{\label{sec:model}Model system}

For the description of ultracold atoms inside a double-well potential, a many-particle Hamiltonian in two-mode approximation
\cite{MilburnEtAl97, *MicheliEtAl03, *MahmudEtAl05, *DusuelVidal05, *Lee06, *Creffield2007}  originally developed in nuclear physics~\cite{LipkinEtAl65} is used:  
\begin{eqnarray}
\label{eq:H}
\hat{H} &=& -\frac{\hbar\Omega}2\left(\hat{c}_1^{\dag}\hat{c}_2^{\phantom\dag}+\hat{c}_2^{\dag}\hat{c}_1^{\phantom\dag} \right) + \hbar\kappa\left(\hat{c}_1^{\dag}\hat{c}_1^{\dag}\hat{c}_1^{\phantom\dag}\hat{c}_1^{\phantom\dag}+\hat{c}_2^{\dag}\hat{c}_2^{\dag}\hat{c}_2^{\phantom\dag}\hat{c}_2^{\phantom\dag}\right)\nonumber\\
&+&\hbar\big(\mu_0+\mu_1\sin(\omega t)\big)\left(\hat{c}_2^{\dag}\hat{c}_2^{\phantom\dag}-\hat{c}_1^{\dag}\hat{c}_1^{\phantom\dag}\right)\;.
\end{eqnarray}
The operator $\hat{c}^{(\dag)}_j$ annihilates (creates) a boson in well~$j$;
$\hbar\Omega$ is the tunneling splitting, $\hbar\mu_0$ is
the tilt between well~1 and well~2 and $\hbar\mu_1$ is the driving amplitude. The interaction
between a pair of particles in the same well is denoted by $2\hbar\kappa$. Dynamics beyond this model has been investigated, e.g., by Refs.~\cite{TrimbornEtAl09, *SakmannEtAl09, *ZollnerEtAl08}.

In order to describe the time-evolution of the inter\-acting system, the Fock basis
$
   |\nu\rangle \equiv |N-\nu,\nu\rangle 
$   
is used. The label   
$   
   \nu=0\ldots N
$
refers to a state with $N-\nu$~particles in well~$1$,
and $\nu$~particles in well~$2$. The Hamiltonian~(\ref{eq:H}) now is the 
sum of two $(N+1)\times(N+1)$-matrices,
\begin{equation}
\label{eq:Hsum}
   H = H_0(t) + H_1 \; .
\end{equation}
While the non-diagonal matrix~$H_1$ is given by the tunneling-terms of Eq.~(\ref{eq:H}), 
the diagonal matrix~$H_0$ includes both the interaction between 
the particles and the applied potential difference. For the solution of the Schr\"odinger equation
the ansatz based on the interaction picture~\footnote{{In the textbook version of the interaction picture~\cite{CohenTannoudjiDiuLaloe} the exponent contains an operator rather than a function. Furthermore, the integral would be the integral from 0 to $t$. While the first point can be derived by projecting the wave-function on the Fock states, to verify that our choice of the integral also leads to correct results can be done by, e.g., treating it as an ansatz.}}
\begin{equation}
   \langle \nu|\psi(t)\rangle = 
   \f_\nu(t)\exp\left[-\frac{\ri}{\hbar}\int
   \langle \nu | H_0(t)|\nu\rangle d t\right] \; 
\end{equation}
turned out to be useful~\cite{WeissJinasundera05}.

For the time-dependent amplitudes~$ \f_\nu(t)$, appendix~\ref{sec:appendixmodel} derives differential equations which are mathematically equivalent to the
$N$-particle Schr\"odinger equation governed by the Hamiltonian~(\ref{eq:H}):
\begin{eqnarray}
\label{eq:wichtigN}
i \dot{a_j}(t) &=& -\frac{\sqrt{N-j+1}\sqrt{j}}{\sqrt{2}}\sum_kA^*_ke^{i\eta^{(j)}_kt} a_{j-1}(t)\nonumber\\
&&-\frac{\sqrt{N-j}\sqrt{j+1}}{\sqrt{2}}\sum_kA_ke^{-i\eta^{(j+1)}_kt} a_{j+1}(t)\;.
\end{eqnarray}
Here, we use the definitions $a_{-1}\equiv 0$ and  $a_{N+1}\equiv 0$ as well as 
\begin{equation}
\label{eq:eta}
\eta_{k}^{(j)}\equiv-k\omega+2\mu_0-2[N-(2j-1)]\kappa\;, \quad j=1\ldots N
\end{equation}
and 
\begin{equation}
\label{eq:defA}
A_j\equiv i^j\frac{1}{\sqrt{2}}\Omega J_j(2\mu_1/\omega)\;,
\end{equation}
where $J_j$ is a Bessel function of integer order (cf.\ appendix~\ref{sec:appendixmodel}).

The Eqs.~(\ref{eq:wichtigN}) and (\ref{eq:eta}) include the experimentally relevant~\cite{CheinetEtAl08} case of $N=2$, for which we introduce the abbreviations $\sigma_k=\eta_k^{(1)}$ and  $\widetilde{\sigma_{\ell}}=\eta_{\ell}^{(2)}$ and thus
 \begin{eqnarray}
\label{eq:sigma}
\sigma_k&\equiv& -k\omega+2\mu_0-2\kappa\nonumber\\
\widetilde{\sigma}_\ell&\equiv& -\ell\omega+2\mu_0+2\kappa\;.
\end{eqnarray}

 An important observable to detect photon-assisted tunneling is the time averaged transfer, 
\begin{equation}
\label{eq:transfer}
\left\langle P_{\rm trans}\right\rangle_T=\frac{1}{NT}\int_0^T{\left\langle \Psi\left(t\right)\left|\hat{c}_2^{\dag}\hat{c}_2^{\phantom\dag}\right|\Psi\left(t\right)\right\rangle}dt\;,
\end{equation}
which is accessible to experimental measurements and describes the expectation value of the time averaged fraction of particles found in the upper well, if all particles are initially prepared in the lower one.

\begin{figure}
\includegraphics[width=0.5\linewidth,angle=-90]{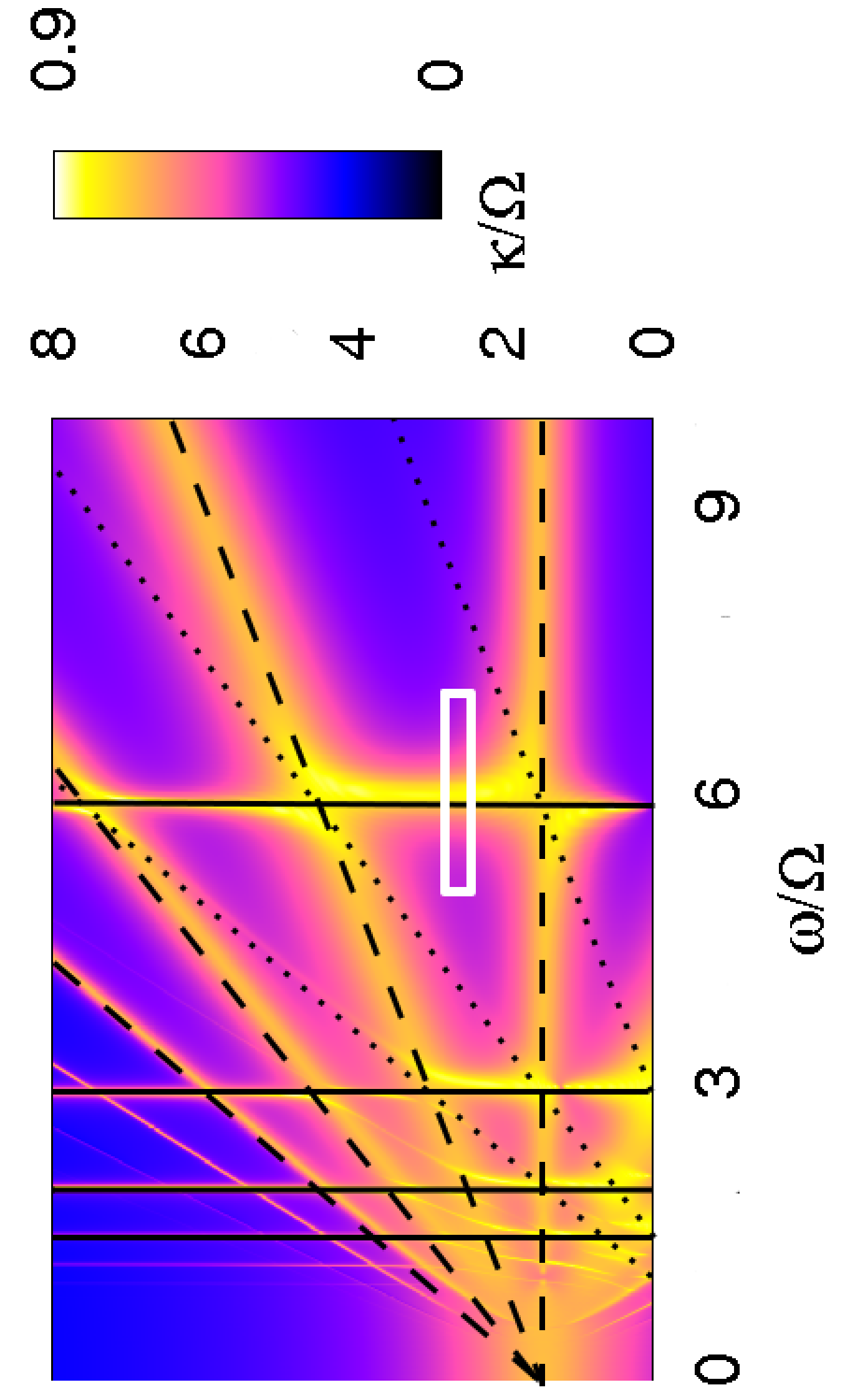}
\caption{\label{fig:N2} (Color online) Two dimensional projection of the (within the model~(\ref{eq:H})) exact particle transfer~(\ref{eq:transfer}) for the experimentally realistic situation~\cite{CheinetEtAl08} of $N=2$ particles in a double well as a function of both shaking frequency~$\omega$ and interaction $\kappa$. The transfer is averaged over $T\Omega=100$ with $2\mu_1/\omega=1.8$, $\mu_0=1.5\Omega$. All particles are initially in the lower well. For better visibility the fourth root of the transfer is plotted; bright colors correspond to large transfer. Solid lines: fractional and integer "photon" resonance, dashed/dotted lines: resonances between the interaction and the tilt plus assistance of photons. White rectangle marks parameter range for the quasi-energy plot in Fig.~\ref{fig:quasi}. The $1/2$ photon resonance is visible as a large effect in the particle transfer near the vertical line at $\omega = 6\Omega$.}
\end{figure}

In Fig.~\ref{fig:N2} a two dimensional projection of the (within the model~(\ref{eq:H})) exact time-averaged particle transfer is shown as a function of the driving frequency $\omega/\Omega$ and the interaction $\kappa/\Omega$ for $N=2$. The dynamics display rich features: The solid lines mark integer and fractional "photon" resonances as discussed in~\cite{TeichmannEtAl09}. This means the tilt $2\hbar\mu_0$ is bridged by a (fractional) integer multiple $m$ of the driving frequency: $m\hbar\omega=2\hbar\mu_0$. The dashed and dotted lines mark resonances between the interaction $\kappa$ and the tilt $\mu_0$ plus the assistance of an integer number of "photons" $k$ and $\ell$ respectively, i.e.\ $\sigma_k=0$ (dashed) and $\widetilde{\sigma}_\ell=0$ (dotted).

Another very important approach towards resonances in a system governed by a Hamiltonian periodic in time are quasi-energies~\nocite{Shirley65}\cite{Shirley65,Zeldovich67}. They account for the problem that for such systems neither energy nor momentum is a reasonable quantum number anymore. According to the Floquet theorem, which is similar to the Bloch theorem known from solid state physics~\cite{AshcroftMermin}, the eigenstates $\Psi(t)$ for a Hamiltonian $H(t+T)\equiv H(t)$ have the form $\Psi(t)=\exp(i\varepsilon t/\hbar)u(t)$ with the function $u(t+T)=u(t)$. The eigenvalues $\varepsilon$ of the new Hamiltonian $\bar{H}=H(t)-i\hbar\partial/\partial t$ are called quasi-energies with the corresponding eigenstates $u(t)$. However, for every quasi-energy there can be found a whole set of further eigenvalues $\varepsilon+m\hbar\omega$ of $\bar{H}$. Therefore the quasi-energies are not unique as the choice of $m\in\mathbb{Z}$ is free, but this has no influence on the physical state.

\begin{figure}
\includegraphics[width=0.5\linewidth,angle=-90]{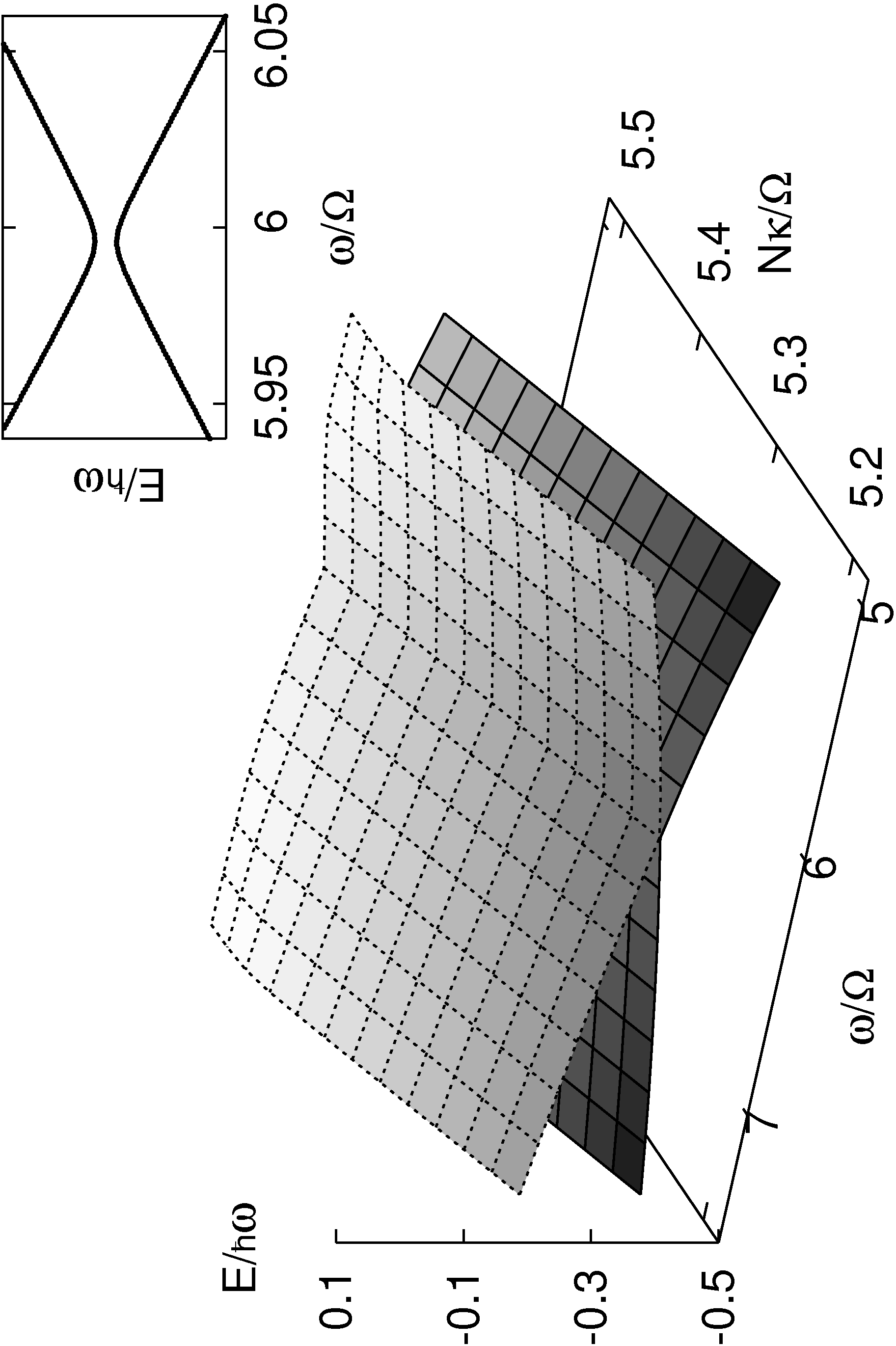}
\caption{\label{fig:quasi} Two of the three quasi-energies for $N=2$ particles. All parameters as in Fig.~\ref{fig:N2} The inset shows these quasi-energies for $N\kappa/\Omega=5.3$. The avoided crossing indicates the $1/2$ "photon" resonance at $\omega/\Omega \simeq 6$.}
\end{figure}
It has been shown~\cite{Shirley65} that avoided crossings in the quasi-energy spectrum correspond to resonances in the transition probability. Thus, in the case discussed here avoided crossings for the right choice of parameters would clearly indicate resonances and support the conclusions already drawn from the map. In the two mode system with $N$ particles there are $N+1$ quasi-energies possible in the range between $n\hbar\omega$ and $(n+1)\hbar\omega$, each corresponding to a certain physical state. In Fig.~\ref{fig:quasi} two of the three quasi-energies for $N=2$ have been plotted. In fact, one recognizes an avoided crossing for the same set of parameters for which the $1/2$ "photon" resonance is observed in Fig.~\ref{fig:N2}.

\begin{figure}
\includegraphics[width=1.0\linewidth]{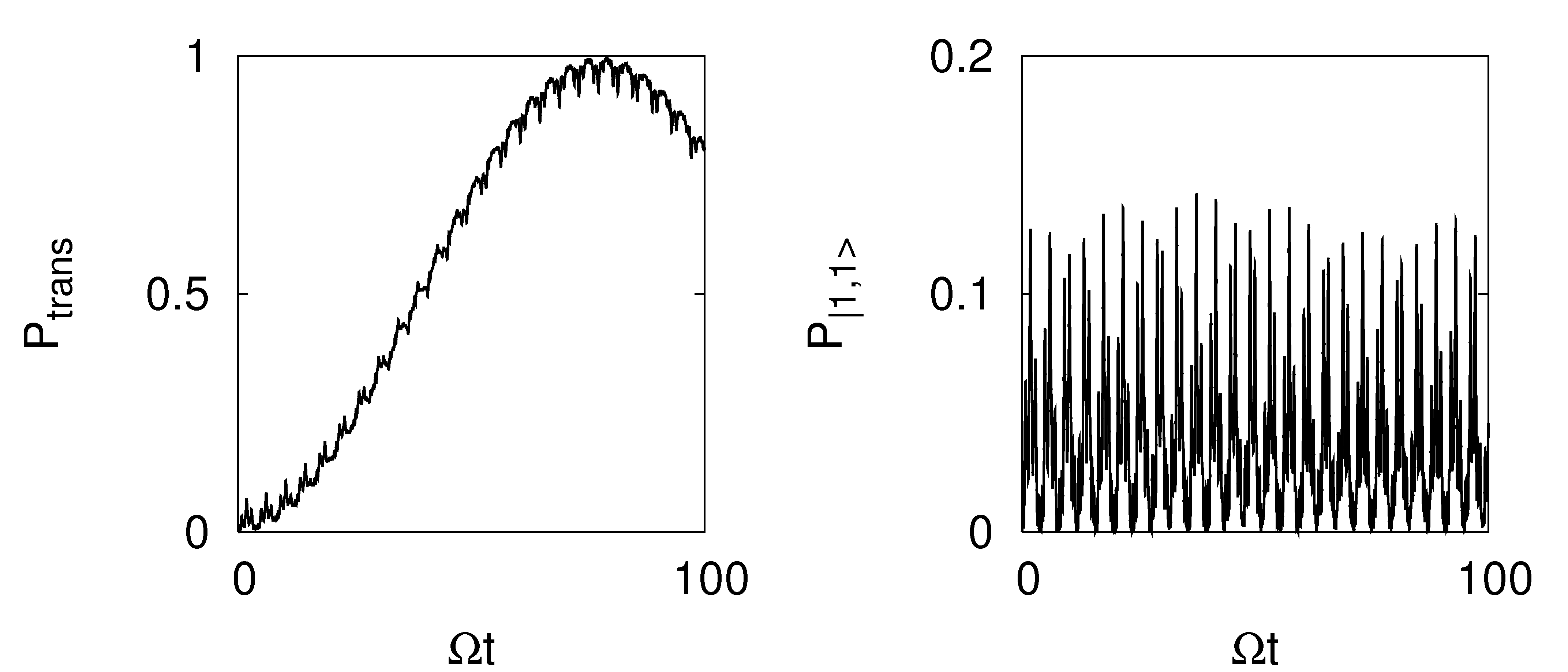}
\caption{\label{fig:NotSTIRAP} 1/2-photon resonance for $N=2$ particles. Left: The nearly perfect transfer between the states occurs for the parameters $\mu_0=1.5\Omega$, $\omega=6\Omega$, $\kappa=0.75\Omega$, $2\mu_1/\omega=3.113$. The transfer was obtained by periodically shaking the double well. Right: The probability to occupy the intermediate state $|1,1\rangle$ through which the transfer takes place.}
\end{figure}
Figure~\ref{fig:NotSTIRAP} demonstrates that fractional photon-assisted tunneling, far from being the small effect of Refs.~\cite{EckardtEtAl05,TeichmannEtAl09,HallerEtAl10,XieEtAl10}, can in fact be a large effect: the transfer between both wells is nearly complete. The tunneling of experimentally realistic particle numbers of $N=2$  from the left to the right well occurs via occupying the intermediate state $|1,1\rangle$. While the probability is small as has to be expected for interacting particles~\cite{Folling07}, it is finite and thus responsible for the transfer. Contrary to the case of STIRAP which avoids occupying unstable intermediate atomic states by depopulating them before populating them by using two laser pulses at different frequencies and time-dependence~\cite{BergmannEtAl98}, there is no need to avoid the population of the intermediate level in our case. Furthermore, only a single shaking frequency is involved. Figure~\ref{fig:NotSTIRAP} also shows that the half-integer resonance is related to the co-tunneling of the repulsively bound pairs of Ref.~\cite{WinklerEtAl06}.

\section{\label{sec:method}Effective model}

The calculations presented in this section show, how an approximation similar to the rotating wave approximation~\cite{HarocheRaimond06} leads from the set of differential equations~(\ref{eq:wichtigN}) to an eigenvalue problem, which may, for instance, be easily solved via numerical diagonalization, even for higher particle numbers. For some special cases such as depicted in appendix~\ref{sec:analytic} the eigenvalues may also be calculated analytically for $N=2$. As higher particle numbers are treated analogously, the calculations in this section are initially performed for $N=2$ particles before giving the general $N$-particle equivalent. Furthermore, the initial state of the system under consideration is always $\left|0\right\rangle$.

While in~\cite{TeichmannEtAl09} time-dependent perturbation theory was applied to an equation equivalent to Eq.~(\ref{eq:wichtigN}), we here use a different approach: large oscillation frequencies are discarded; this occurs in each of the sums in Eq.~(\ref{eq:matrix1}) - which is the 2-particle version of Eq.~(\ref{eq:wichtigN}) - only the phase factor with the smallest frequency $\sigma_{k'}$, and $\widetilde{\sigma}_{\ell'}$ respectively, is maintained, according to the definition
\begin{eqnarray}
\label{eq:sigk}
\lambda&\equiv&\sigma_{k'}\\
\label{eq:sigell}
\nu&\equiv&\widetilde{\sigma}_{\ell'}\;,
\end{eqnarray}
where $k'$ is the integer for which $\left\{\left|-k\omega+2\mu_0-2\kappa\right|;k\in\mathbb{Z}\right\}$ reaches its minimum and, analogously, ${\ell'}$ is the integer for which
$\left\{\left|-\ell\omega+2\mu_0+2\kappa\right|;\ell\in\mathbb{Z}\right\}$ reaches its minimum.

One now has to solve the differential equations
\begin{equation}
\label{eq:matrix2}
i\left(
\begin{array}{c}
\dot{a_0}(t)\\
\dot{a_1}(t)\\
\dot{a_2}(t)
\end{array}\right)=
\left(\begin{array}{c}
0 \quad -A_{k'}e^{-i\lambda t} \quad 0\\
-A^*_{k'}e^{i\lambda t} \quad 0 \quad -A_{\ell'}e^{-i\nu t}\\
0 \quad -A^*_{\ell'}e^{i\nu t} \quad 0
\end{array}\right)
\left(\begin{array}{c}
a_0(t)\\
a_1(t)\\
a_2(t)
\end{array}\right).
\end{equation}

The ansatz
\begin{eqnarray}
\label{eq:ansatz}
a_0(t) &=& \widetilde{a}_0 e^{-i\omega t}\;,\nonumber\\
a_1(t) &=& \widetilde{a}_1 e^{-i(\omega-\lambda)t}\;,\nonumber\\
a_2(t) &=& \widetilde{a}_2 e^{-i(\omega-\lambda-\nu)t}
\end{eqnarray}
yields a time-independent eigenvalue problem
\begin{equation}
\label{eq:matrix3}
\omega\left(
\begin{array}{c}
\widetilde{a}_0\\
\widetilde{a}_1\\
\widetilde{a}_2
\end{array}\right)=\mathbf B
\left(\begin{array}{c}
\widetilde{a}_0\\
\widetilde{a}_1\\
\widetilde{a}_2
\end{array}\right)
\end{equation}
with
\begin{equation}
\label{eq:matb}
\mathbf B = \left(\begin{array}{lcr}
0 & -A_{k'} & 0\\
-A^*_{k'} & \lambda & -A_{\ell'}\\
0 & -A^*_{\ell'} & \lambda+\nu
\end{array}\right)
\end{equation}
which yields three linearly independent solutions of the type~(\ref{eq:ansatz}). Note that the amplitudes $a_j(t)$ are functions of time, while the amplitudes $\widetilde{a}_j$ of the eigenvalue problem are not; furthermore, as soon as three linearly independent solutions are found this also implies that one knows all solutions of the linear first-order differential equation~(\ref{eq:matrix2}).
The $N$-particle equivalent of Eqs.~(\ref{eq:ansatz}) and (\ref{eq:matrix3}) can be found in appendix~\ref{app:effectN}.

Thus, by choosing the smallest frequencies as suggested by the rotating wave approximation, we arrived at a time-independent equation which tells us via minimizing the modulus of Eqs.~(\ref{eq:sigk}) and (\ref{eq:sigell}) and inserting the resulting integers into Eq.~(\ref{eq:defA}) which Bessel functions will be relevant for the tunneling dynamics. As we show in the following, the effective model (\ref{eq:ansatz})-(\ref{eq:matb}) (and its $N$-particle version, see Eqs.~(\ref{eq:eigen})-(\ref{eq:loesung})) well describes the exact numerics and it is thus a tool to predict interesting parameter regimes for future experiments. At the same time, it offers a much simpler approach to understanding fractional photon-assisted tunneling than previous research~\cite{TeichmannEtAl09,XieEtAl10}. As expected, for vanishing interactions Eq.~(\ref{eq:eta}) shows that only a single Bessel function remains thus leading to, e.g.,  the integer photon-assisted tunneling investigated in Refs.~\cite{EckardtEtAl05,SiasEtAl08}. In general, more than one Bessel function will be relevant in a much simpler way than predicted in the approach of Ref.~\cite{TeichmannEtAl09}.

However, before testing the effective model~(\ref{eq:eigen})-(\ref{eq:loesung}), it should be noted that the choice of the smallest frequency might not be uniquely defined: rather than minimizing each $|\eta^{(j)}_k|$, it might in fact be preferable to minimize the modulus of the entries on the diagonal of $\mathbf B$, starting from the top. For the $N=2$-particle case one thus has:
\begin{equation}
\label{eq:sigk2}
\nu = \widetilde{\sigma}_{\ell''}\;,
\end{equation}
where $\ell''$ is the integer $\ell$ for which 
$\left\{\left|(k'+\ell)\omega-4\mu_0\right|;\ell\in\mathbb{Z}\right\}$
reaches its minimum [where $k'$ is the integer defined below Eq.~(\ref{eq:sigk})].
 This second approach (to minimize the sum) will also be tested in this paper.

\section{\label{sec:results}Results}
The effective model allows us to understand the physics both via analytic calculations and by exact numeric diagonalisation.
We start with the condition $\omega/2=2\mu_0$, corresponding to the case when the tilt between the individual wells is bridged by half the photon energy. This implies $\sigma_k=-\omega(k-1/2)-2\kappa$ and $\widetilde{\sigma}_\ell=-\omega(\ell-1/2)+2\kappa$. Now the condition $\sigma_{k'}+\widetilde{\sigma}_{\ell'}\stackrel{!}{=}0$, which corresponds to the half integer resonance~\cite{TeichmannEtAl09}, results in $k'+\ell'=1$ (cf.\ Eqs~(\ref{eq:sigk}) and (\ref{eq:sigell})), thus leading to the eigenvalue problem
\begin{equation}
\label{eq:matrix4}
\omega\left(
\begin{array}{c}
\widetilde{a}_0\\
\widetilde{a}_1\\
\widetilde{a}_2
\end{array}\right)=
\left(\begin{array}{lcr}
0 & -i^{k'}\Omega_1 & 0\\
-i^{-k'}\Omega_1 & \sigma_{k'} & -i^{1-k'}\Omega_2\\
0 & -i^{k'-1}\Omega_2 & 0
\end{array}\right)
\left(\begin{array}{c}
\widetilde{a}_0\\
\widetilde{a}_1\\
\widetilde{a}_2
\end{array}\right).
\end{equation}
Here the following definition for the tunneling frequencies $\Omega$ has been adopted:
\begin{eqnarray}
\Omega_1&=&\frac{\Omega}{\sqrt{2}}J_{k'}\left(\frac{2\mu_1}{\omega}\right)\;,\nonumber\\
\Omega_2&=&\frac{\Omega}{\sqrt{2}}J_{1-k'}\left(\frac{2\mu_1}{\omega}\right)\;.
\label{eq:Omega12}
\end{eqnarray}
Analytic solutions are derived in the appendix~\ref{sec:analytic}.

In Fig.~\ref{fig:transfer} both the analytic expression and the (within the model~(\ref{eq:H})) exact numerical solution are displayed. Note that furthermore the parameters are chosen such that  $|J_0(2\mu_1/\omega)|=|J_1(2\mu_1/\omega)|$. The approximation shows very good agreement with the numerics; especially for values close to $\kappa=\omega/4$ the relative error, plotted in the inset, almost becomes zero. In experiments like~\cite{CheinetEtAl08} and under conditions such that the single band approximation used in Eq.~(\ref{eq:H}) is valid, $\kappa/\Omega$ can experimentally be determined with at least 1\% accuracy \cite{Trotzky10}.  Thus the full width at half maximum obtained from our calculations is sufficiently large to be very promising for further experiments. 

\begin{figure}
\includegraphics[width=0.6\linewidth,angle=-90]{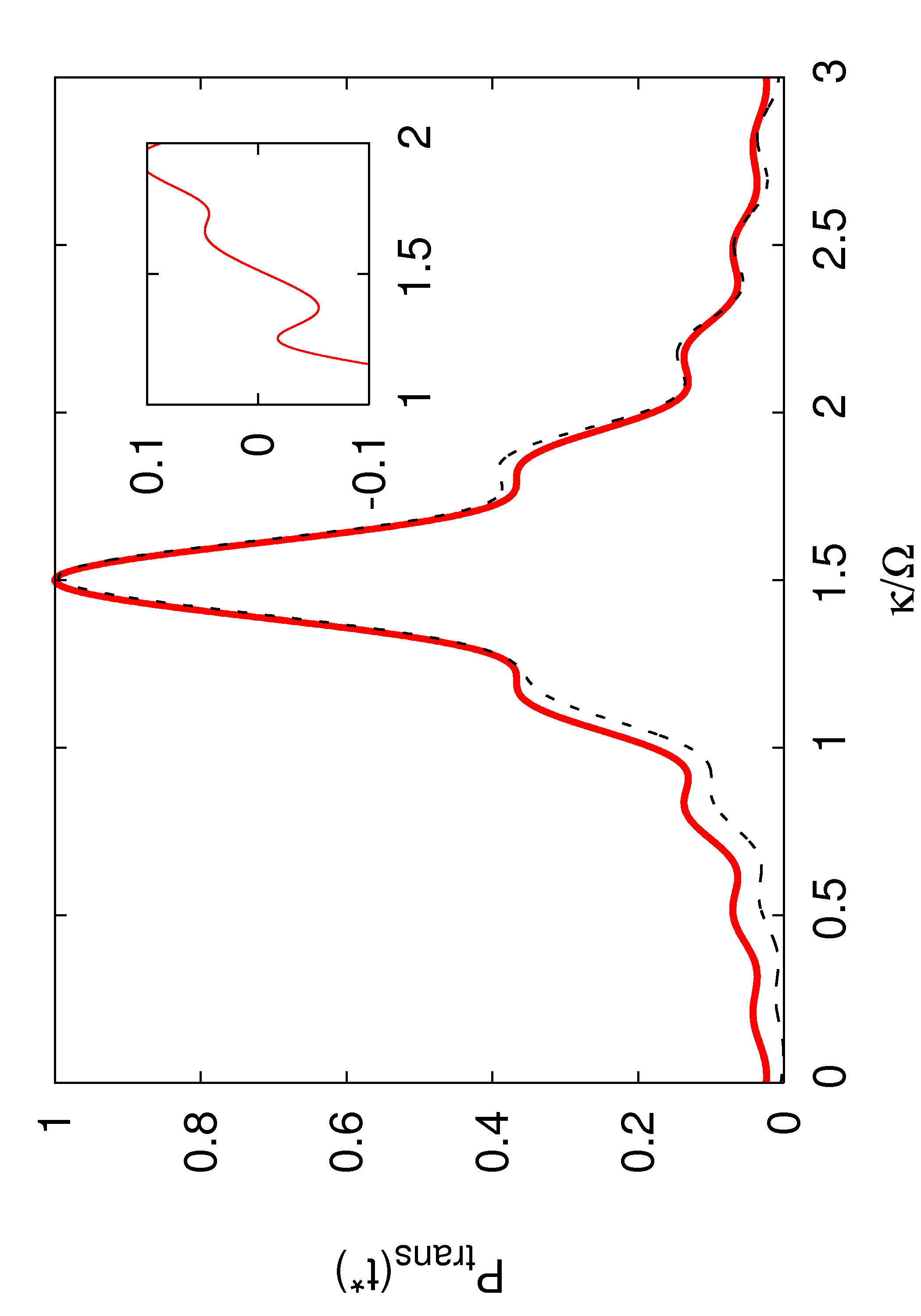}
\caption{\label{fig:transfer} (Color online) Time dependent transfer $P_{\rm trans}$~(\ref{eq:transfer}) for $N=2$ particles with $2\mu_1/\omega=3.113$, $\mu_0=\omega/4$ and $t^*$ denotes the time for which perfect
transfer is suggested from the analytical calculations (cf.\ Eq.~(\ref{eq:tstar})). We start with the experimentally realistic~\cite{Foelling10} initial condition that all particles are in the lower well (cf.\ Ref.~\cite{CheinetEtAl08}). Solid line: exact numerics within the model~(\ref{eq:H}), dashed line: analytical approximation obtained by inserting Eq.~(\ref{eq:tstar}) into Eq.~(\ref{eq:imbalance3}). The inset displays the relative error.}
\end{figure}

\begin{figure}
\includegraphics[width=0.5\linewidth,angle=-90]{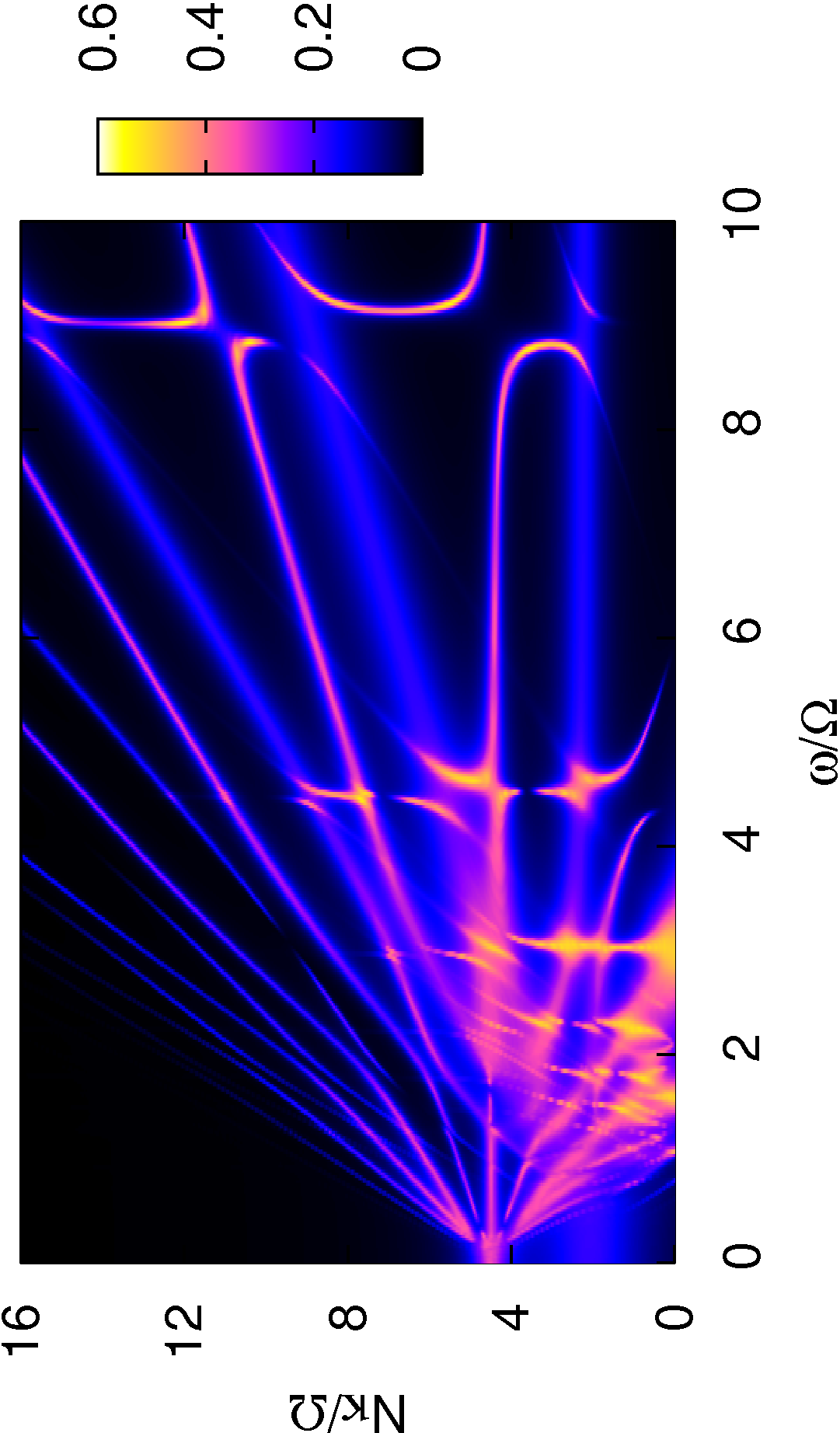}
\caption{\label{fig:exact} (Color online) Two dimensional projection of the (within the model~(\ref{eq:H})) exact particle transfer for $N=3$ particles averaged over $T\Omega=100$ with $2\mu_1/\omega=1.8$, $\mu_0=1.5\Omega$. All particles are initially in the lower well of one of the double-well potentials of Ref.~\cite{CheinetEtAl08}. The $1/3$-photon resonance is a large effect visible in the particle transfer near the vertical line at $\omega = 9\Omega$.}
\end{figure}

Figure~\ref{fig:exact} shows a map in which the (within the model~(\ref{eq:H})) exact time averaged particle transfer~(\ref{eq:transfer}) is displayed for $N=3$ particles as a function of the coupling strength $N\kappa/\Omega$ and the driving frequency $\omega/\Omega$. Again we see many of the features already obtained from the two particle solution. For two particles, a half-integer resonance is visible as a straight vertical line at $\omega = 6\Omega$ in Fig.~\ref{fig:N2} whereas in the three-particles plot Fig.~\ref{fig:exact} only a resonance moving for increasing interaction to the left of this frequency is observable (in agreement with Ref.~\cite{LuEtAl10}). 

\begin{figure}
\includegraphics[width=5cm,angle=-90]{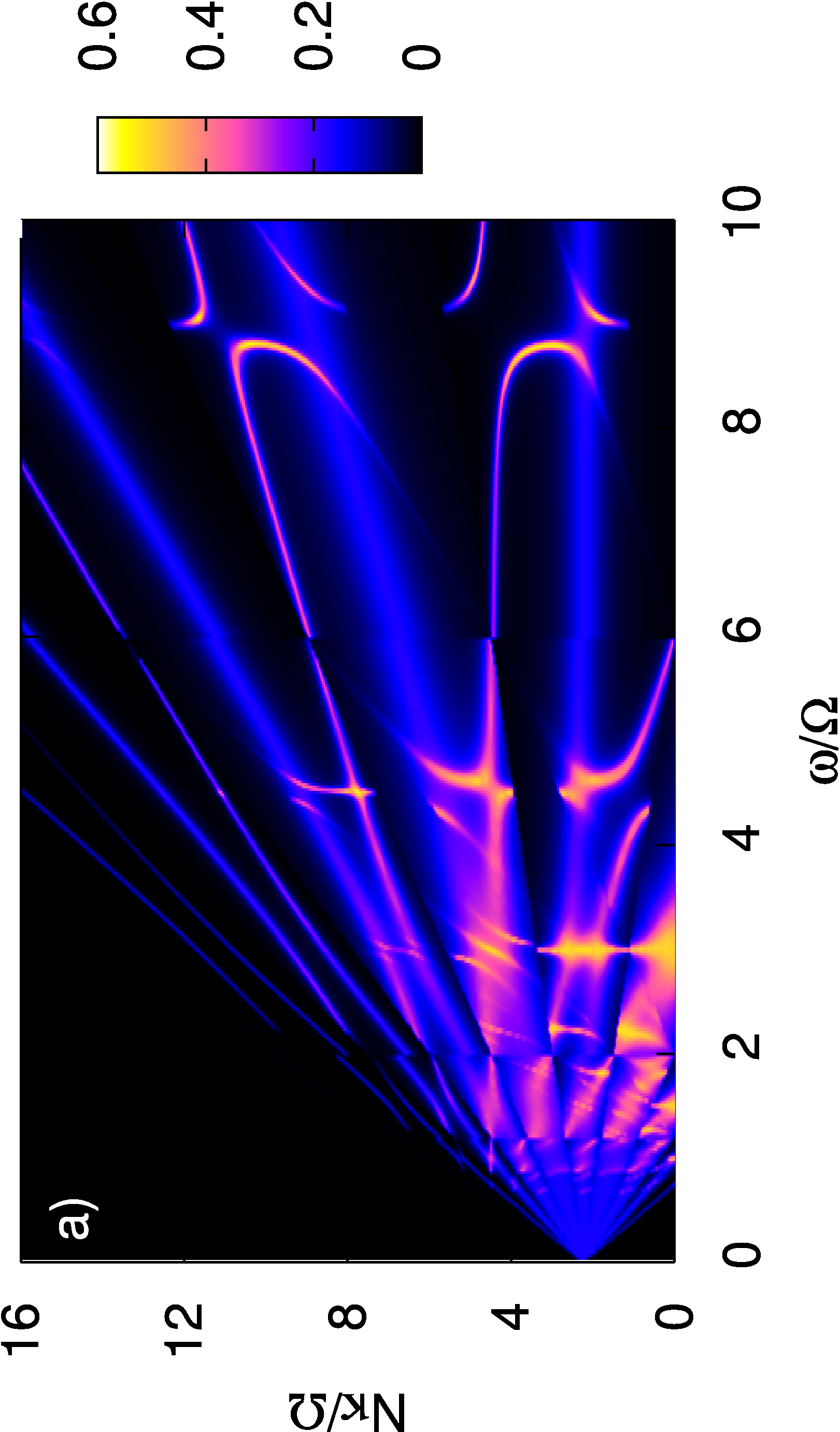}
\includegraphics[width=5cm,angle=-90]{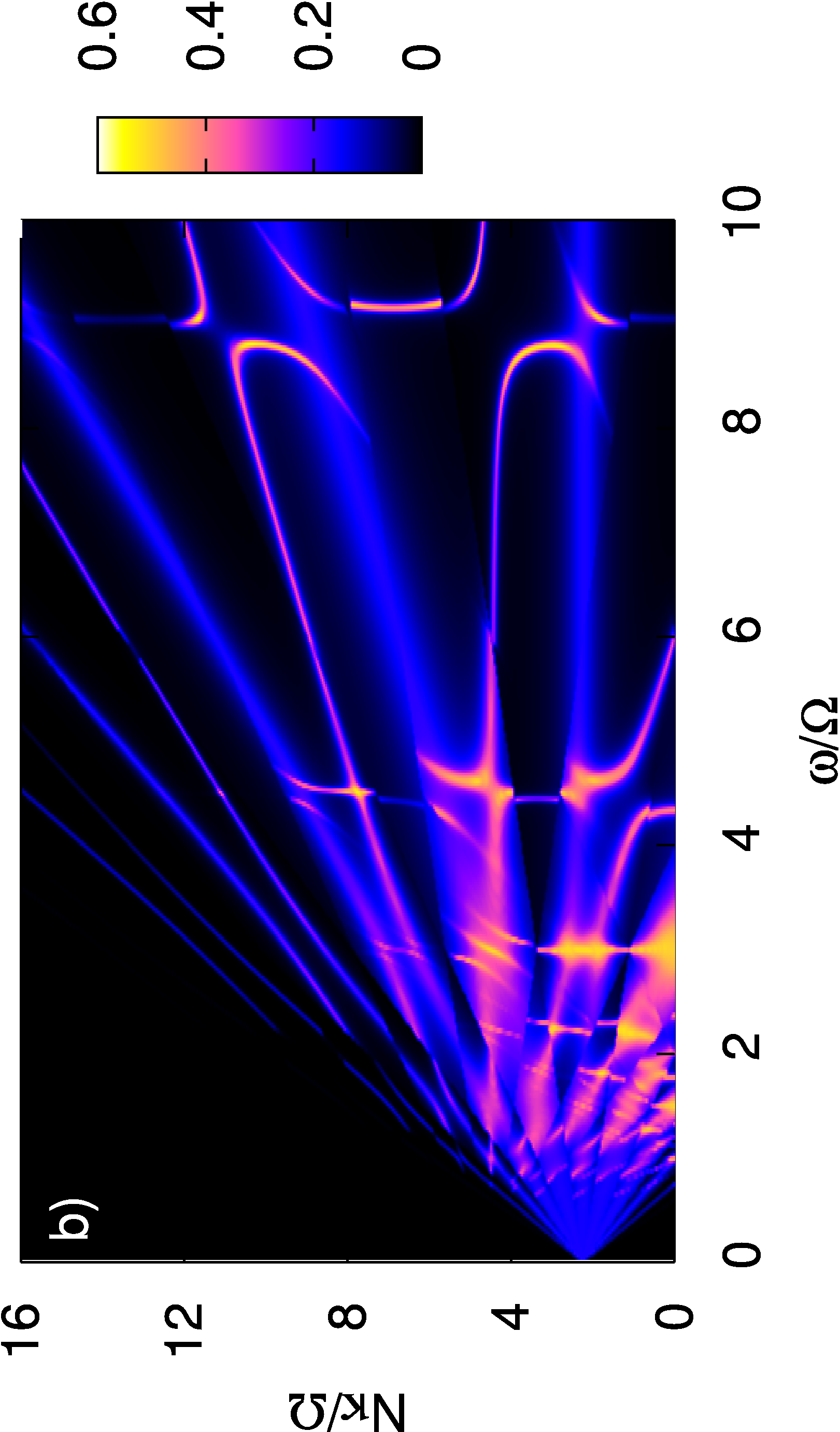}
\includegraphics[width=5cm,angle=-90]{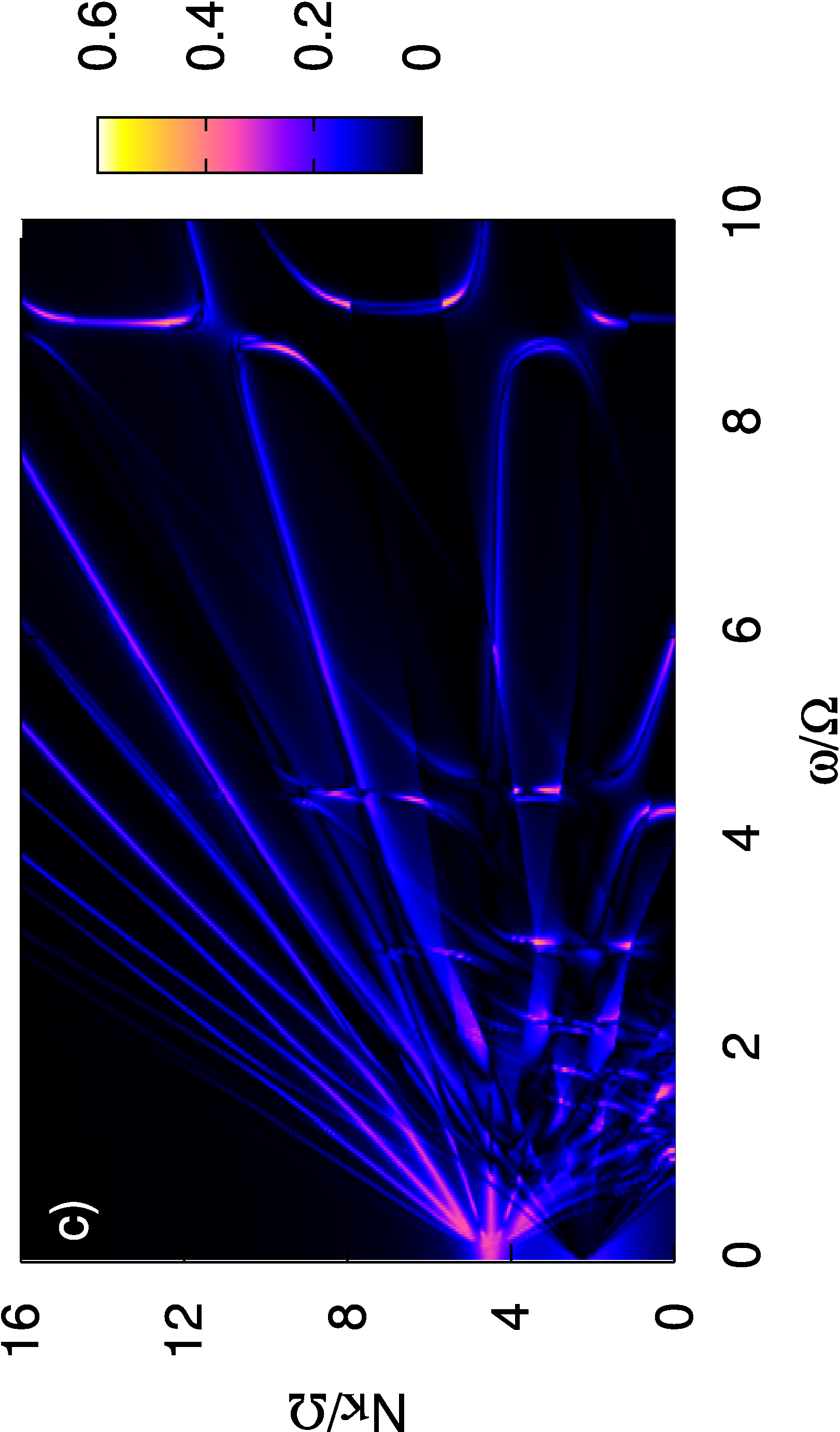}
\caption{\label{fig:analytic} (Color online) Two dimensional projections of the time averaged particle transfer. All parameters as in Fig.~\ref{fig:exact}. (a) Approximated transfer as given by the model~(\ref{eq:matrix3N}) for $N=3$ particles by minimizing $|\eta^{(j)}_k|$ for all $j=1,2,3$ separately. (b) Approximated transfer for $N=3$ particles by minimizing the sums $|\sum_{\ell=1}^{j}\eta^{(j)}_k|$  for $j=1,2,3$ starting with low $j$. (c) Modulus of the difference of the approximated transfer displayed in (b) and the exact solution from Fig.~\ref{fig:exact}.}
\end{figure}

For the same parameters as in Fig.~\ref{fig:exact} the results for the two approximations have been plotted in Figs.~\ref{fig:analytic} (a) and (b), as well as the difference between the second approximation and the exact solution, which is plotted in Fig.~\ref{fig:analytic} (c).

\begin{figure}
\includegraphics[width=0.7\linewidth]{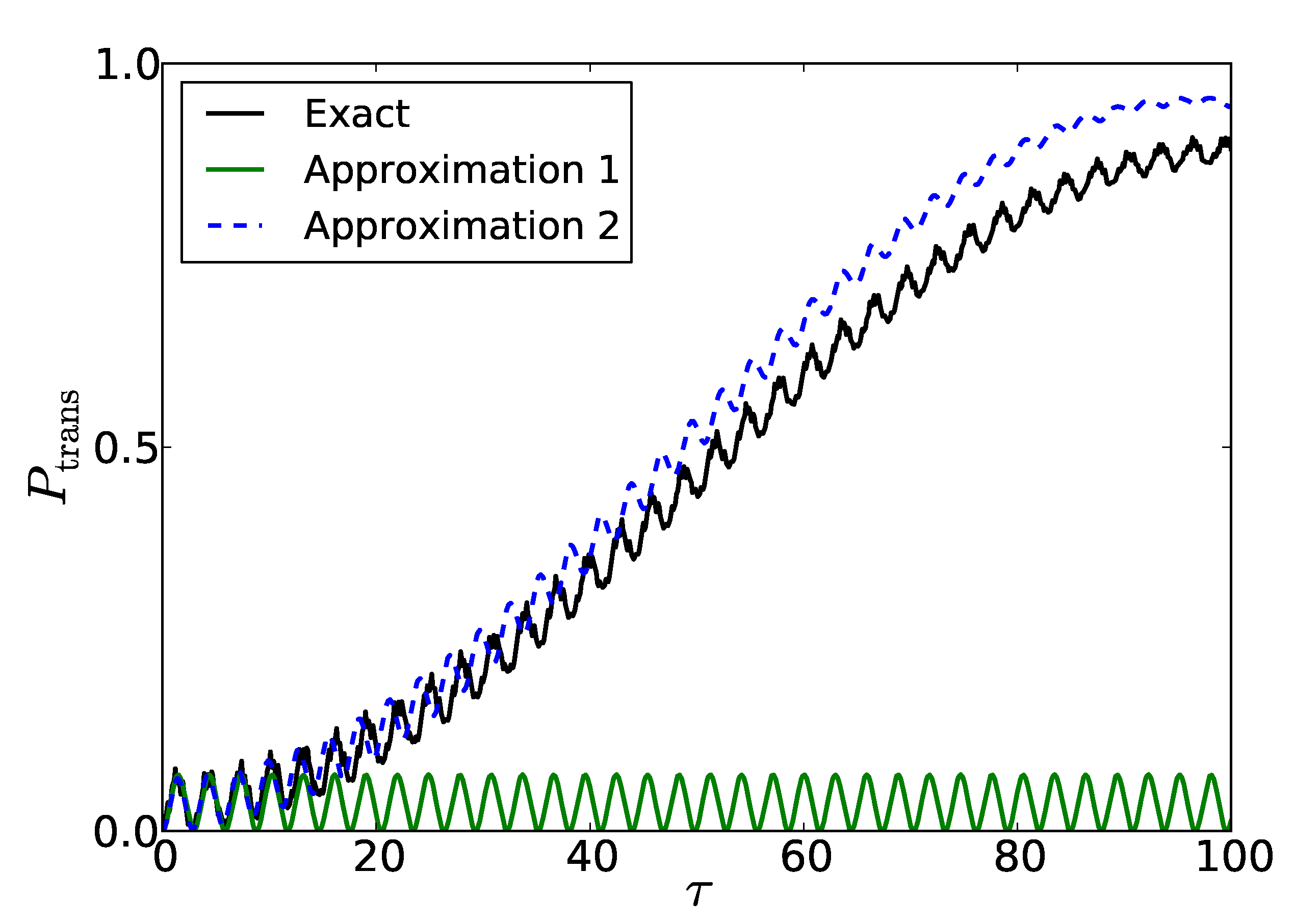}
\caption{\label{fig:time} (Color online) Time dependent particle transfer over dimensionless time $\tau$ for $N=3$, $2\mu_0/\Omega=3$, $2\mu_1/\omega=1.8$, $\omega/\Omega=9.175$, $N\kappa/\Omega=7.68$. solid (black) line: exact numerics, solid gray (green) line: Eq.~(\ref{eq:matrix3N}) with  minimizing each $|\eta^{(j)}_k|$ separately for $j=1,2,3$, dashed (blue) line:  Eq.~(\ref{eq:matrix3N}) with minimizing the sums  $|\sum_{\ell=1}^{j}\eta^{(j)}_k|$  for $j=1,2,3$ starting again from low $j$.}
\end{figure}

It is clearly visible that both approximations based on Eq.~(\ref{eq:matrix3N}) are fairly good reproductions of the (within the model~(\ref{eq:H})) exact solution. However, in the case analyzed here the second approach, in which the sums $|\sum_{\ell=1}^{j}\eta^{(j)}_k|$  are minimized for $j=1,2,3$, shows better agreement with the exact numerics. The probably most outstanding feature to see this is the ${1}/{3}$-"photon" resonance at ${\omega}=9{\Omega}$, that partially vanishes in the first approximation and stays more visible in the second. In order to see, whether this feature in fact indicates that the second approximation is for this set of parameters superior to the first, we further analyze the time dependent evolution of the system for $N=3$. 
In Fig.~\ref{fig:time} both approximations and the exact particle transfer are plotted. One clearly recognizes that the second approximation much better reflects the actual behavior of the system contrary to the first. The good agreement between the time-evolutions excludes accidental agreement on the level of the time-averaged dynamics.

Both approximations show good agreement in the relevant parameter regime (depicted by bright colors in the two dimensional-projection plots) for large enough frequencies while they 
do not fit well for very low frequencies. This is because we deal with a high frequency approximation, i.e.\ the lower the driving frequency is, the more relevant the discarded higher frequencies become.

\begin{figure}
\includegraphics[width=0.6\linewidth,angle=-90]{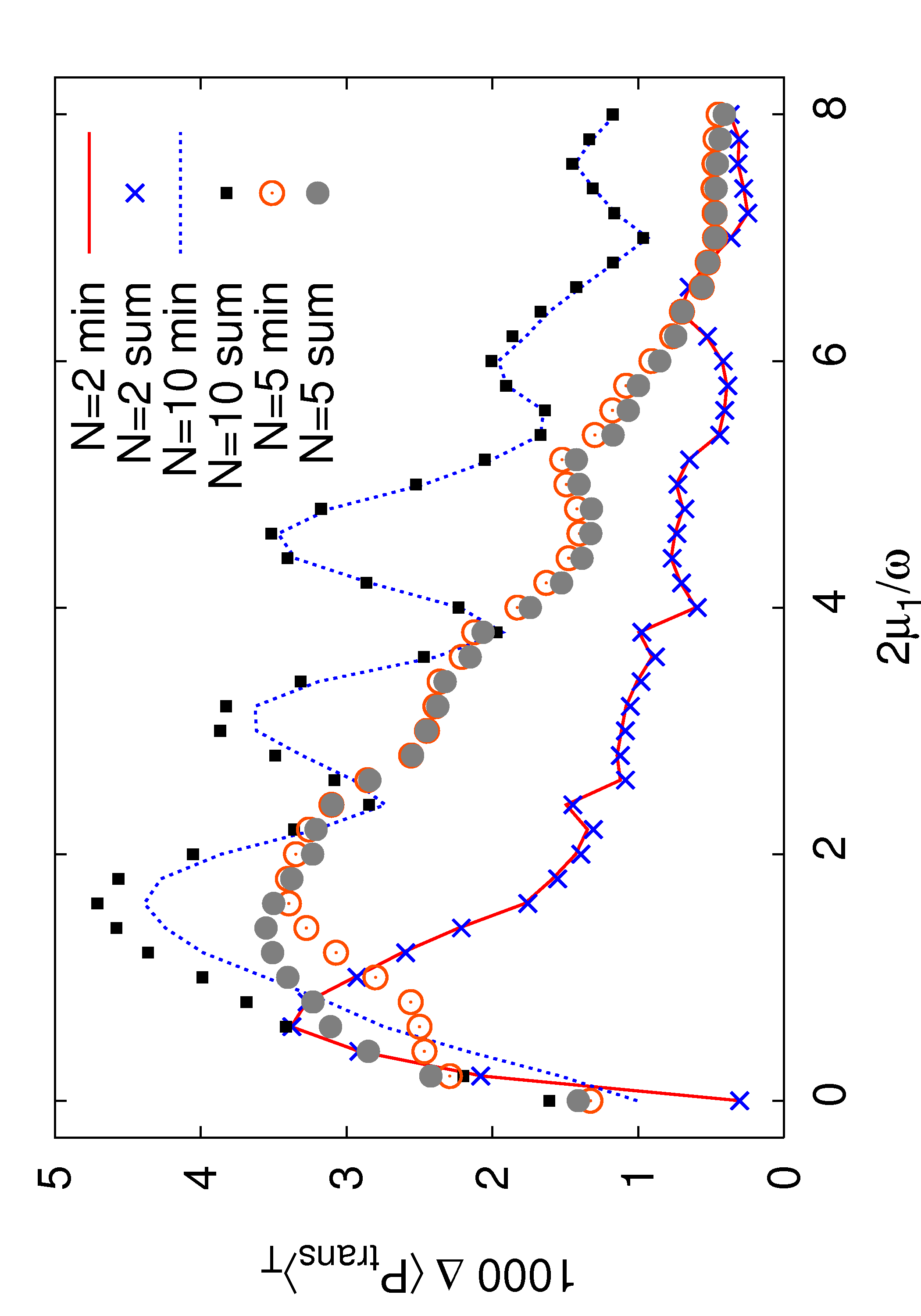}
\caption{\label{fig:comp} (Color online) Comparison of the two different approximations based on Eq.~(\ref{eq:matrix3N}) (min: minimizing $|\eta^{(j)}_k|$ separately for $j=1,2,\ldots, N$, sum: minimizing the sums $|\sum_{\ell=1}^{j}\eta^{(j)}_k|$  for $j=1,2,\ldots, N$ starting again from low $j$) as a function of the driving amplitude $2\mu_1/\omega$ and for different particle numbers which can be realized with the experimental setup of Ref.~\cite{CheinetEtAl08}. Displayed is the sum of the squared deviations from the exact particle transfer across the map~(\ref{eq:deviationsum}).}
\end{figure}
In order to decide which of the two approximations should be preferred for the calculation of the time averaged particle transfer we have further calculated the sums of the squared deviations from the exact particle transfer,
\begin{equation}
\label{eq:deviationsum}
\Delta \left\langle P_{\rm trans}\right\rangle_T=\frac{\int_{3}^{10}d\omega\int_{0}^{16}d(N\kappa)\left|\left\langle P_{\rm trans}\right\rangle_T-\left\langle\right. \widetilde{P}_{\rm trans}\left.\right\rangle_T\right|^2 }{\int_{3}^{10}d\omega\int_{0}^{16}d(N\kappa)1}\;,
\end{equation}
 across the map as a function of the driving amplitude $2\mu_1/\omega$ and for different particle numbers $N=2,5,10$. To exclude low frequencies for which the approximations are both not reliable, the driving frequency ranges from $\omega/\Omega=3$ to $\omega/\Omega=10$ in steps of $0.025$. The interaction strength ranges from $N\kappa/\Omega=0$ to $N\kappa/\Omega=16$ in steps of $0.04$. The results are displayed in Fig.~\ref{fig:comp}. From this comparison one cannot clearly decide in favor of one of the two approximations. Depending on the driving amplitude and particle number each of them yields better results for certain cases, contrary to what was presupposed.

\section{\label{sec:conclusion}Conclusion}

Photon-assisted tunneling was investigated for ultra-cold atoms in a periodically shaken optical superlattice. Experimentally realistic particle numbers of less than 4 particles per double well~\cite{CheinetEtAl08} were chosen which turned out to be ideal to investigate fractional photon-assisted tunneling; the calculations were done  under the experimentally realistic condition that tunneling between neighboring double wells can be discarded~\cite{Foelling10}. The numeric and analytic calculations were partially done in the effective model derived by combining the interaction picture with the rotating wave approximation. While one- or two-photon resonances~\cite{SiasEtAl08} are a single particle effect, fractional photon-resonances are a true many-particle quantum effect which only occurs for interacting particles. Although previous investigations seem to indicate that fractional photon-assisted tunneling is a small higher order effect~\cite{EckardtEtAl05,TeichmannEtAl09,HallerEtAl10,XieEtAl10}, we have demonstrated that for the particle numbers of Ref.~\cite{CheinetEtAl08} both the $1/2$- and the $1/3$-resonance can be large effects visible in the transfer of particles between neighboring wells. For the parameters leading to the complete transfer of two particles, the agreement between the exact numerics and the effective model is astonishingly good. In general, the effective model well describes the physics of the system if the system is close to a resonance describing integer or fractional photon-assisted tunneling.

Unlike the perturbation theory approach of Ref.~\cite{TeichmannEtAl09}, the results of the present manuscript are neither restricted to short time-scales nor to fractional photon-assisted tunneling being a small effect. Similar to the rotating wave approximation, higher frequencies were neglected which will, for larger time-scales, eventually lead to deviations even for cases like in Fig.~\ref{fig:transfer} where numerical  and analytical results agree quite well. The transfer via photon-assisted tunneling can be observed with an existing experimental setup~\cite{CheinetEtAl08}. 

\textit{Note added:} Recently, a first experiment with periodically shaken superlattice  has been performed by Chen \textit{et al.}\/ \cite{ChenEtAl11}. The experiment uses the setup of Ref.~\cite{CheinetEtAl08}; photon assisted tunneling is clearly visible both for single particles and for few particles.  One of the results is that the half-integer resonance (labeled ``co-tunneling of a pair of atoms'' by Chen \textit{et al.}, cf.\ Ref.~\cite{WinklerEtAl06}) is clearly visible for the case of two particles per double well. As predicted by the present manuscript, the half-integer resonance does  indeed have a large effect on the particle transfer~\cite{ChenEtAl11}.

\acknowledgments

We thank M.\ Holthaus for his support and S.\ F\"olling, O.\ Morsch and S.\ Trotzky for discussions. Computer power was obtained from the GOLEM I cluster
of the University of Oldenburg.
 ME acknowledges funding by
the Studienstiftung des deutschen Volkes. NT acknowledges funding by the University of Oldenburg. 

\begin{appendix}
\section{\label{sec:appendixmodel}Model}

A set of differential equations was derived~\cite{WeissJinasundera05} which is mathematically equivalent to the
$N$-particle Schr\"odinger equation governed by the Hamiltonian~(\ref{eq:H}):
\begin{eqnarray}
\label{eq:equiv}
   \ri\hbar\dot{\f}_{\nu}(t) & = &
   \langle\nu | H_1 |\nu\!+\!1\rangle\xx_{\nu}(t){\f}_{\nu+1}(t)
\nonumber\\
   & + & \langle\nu|H_1|\nu\!-\!1\rangle \xx_{\nu-1}(t)^*{\f}_{\nu-1}(t) \;.
\end{eqnarray}
In Eq.~(\ref{eq:equiv}), the notation
$
   a_{-1}(t) \equiv a_{N+1}(t) \equiv 0 \, ,
$
was adopted; the phase factors are given by:
\begin{equation}
\label{eq:phase}
   \xx_{\nu}(t) = 
\exp\left({\textstyle \ri\left[2(N - 1 - 2\nu)\kappa t 
   -2\mu_0t+2\mu_1\cos(\omega t)/\omega\right] }\right).
\end{equation}
Furthermore, there are the matrix elements of the tunneling part of the Hamiltonian in the Fock basis:  
\begin{eqnarray}
\langle\nu | H_1 |\nu\!+\!1\rangle &=& \langle N-\nu,\nu | H_1|N-\nu-1,\nu+1\rangle\nonumber\\
&=&  - \frac{\hbar\Omega}2 \sqrt{N-\nu}\sqrt{\nu+1}
\end{eqnarray}
and
\begin{equation}
\langle\nu | H_1 |\nu\!-\!1\rangle = - \frac{\hbar\Omega}2 \sqrt{N-\nu+1}\sqrt{\nu}
\end{equation}

To simplify the expression
for subsequent integrals, we use the expansion in terms of Bessel functions~\cite{Abramowitz84}
\begin{equation}
\label{eq:bessel}
e^{iz\cos(\omega t)}=\sum_{k=-\infty}^{\infty}J_{k}(z)i^ke^{ik\omega t}\,.
\end{equation}

Applying the above expansion and introducing the abbreviation
\begin{equation}
\label{eq:defAap}
A_j=i^j\frac{1}{\sqrt{2}}\Omega J_j(2\mu_1/\omega)
\end{equation}
and the frequencies
\begin{eqnarray}
\label{eq:sigmaApp}
\sigma_k&\equiv& -k\omega+2\mu_0-2\kappa\nonumber\\
\widetilde{\sigma}_\ell&\equiv& -\ell\omega+2\mu_0+2\kappa
\end{eqnarray}
the differential equations~(\ref{eq:equiv}) for a system with $N=2$  particles - the smallest particle number for which the interaction induced fractional photon resonances will be observable with the experimental setup of Ref.~\cite{CheinetEtAl08} - read:
\begin{widetext}
\begin{equation}
\label{eq:matrix1}
i\left(
\begin{array}{c}
\dot{a_0}(t)\\
\dot{a_1}(t)\\
\dot{a_2}(t)
\end{array}\right)=
\left(\begin{array}{lcr}
0 & -\sum_kA_ke^{-i\sigma_kt} & 0\\
-\sum_kA^*_ke^{i\sigma_kt} & 0 & -\sum_\ell A_\ell e^{-i\widetilde{\sigma}_\ell t}\\
0 & -\sum_\ell A^*_\ell e^{i\widetilde{\sigma}_\ell t} & 0\end{array}\right)
\left(\begin{array}{c}
a_0(t)\\
a_1(t)\\
a_2(t)
\end{array}\right)\,.
\end{equation}
The generalization to the $N$-particle-equivalent displayed in Eq.~(\ref{eq:wichtigN}) is straight-forward.

\section{\label{app:effectN}Effective model for $N$ particles}
In order to derive the $N$-particle equivalent of Eqs.~(\ref{eq:ansatz}) and (\ref{eq:matrix3}), we transfer the reasoning involved in the derivation of those equations to the general case: 
let $k_{j}$ be the integer for which $|\eta^{(j)}_k|$ reaches its minimum. Combining the equivalent of Eq.~(\ref{eq:matrix3}) with the definition~(\ref{eq:defA}) leads to an eigenvalue equation,
\begin{equation}
\label{eq:eigen}
\omega\left(
\begin{array}{c}
\widetilde{a}_0\\
\widetilde{a}_1\\
\widetilde{a}_2\\
\ldots\\
\widetilde{a}_{N-1}\\
\widetilde{a}_N
\end{array}\right)={\mathbf B}\left(
\begin{array}{c}
\widetilde{a}_0\\
\widetilde{a}_1\\
\widetilde{a}_2\\
\ldots\\
\widetilde{a}_{N-1}\\
\widetilde{a}_N
\end{array}\right)\;,
\end{equation}
with the tridiagonal Hermitian matrix
\begin{equation}
\label{eq:matrix3N}
\footnotesize
{\mathbf B}\equiv
\left(\begin{array}{cccccc}
0 & -\frac{i^{k_1}\sqrt{N}}{{2}}\Omega J_{k_1}(\frac{2\mu_1}{\omega}) & 0&\ldots&0&0\\
- \frac{i^{-k_1}\sqrt{N}}{{2}}\Omega J_{k_1}(\frac{2\mu_1}{\omega})\!\!\!\!\!\!& \eta^{(1)}_{k_1} &\!\!\!\!\!\! -\frac{i^{k_2}\sqrt{N-1}\sqrt{2}}{{2}}\Omega J_{k_2}(\frac{2\mu_1}{\omega})&\ldots&0&0\\
0 & - \frac{i^{-k_2}\sqrt{N-1}\sqrt{2}}{{2}}\Omega J_{k_2}(\frac{2\mu_1}{\omega})& \eta^{(1)}_{k_1}+ \eta^{(2)}_{k_2}&\ldots&0&0\\
\ldots&\ldots&\ldots&\ldots&\ldots&\ldots\\
0&0&0&\ldots&\sum_{j=1}^{N-1}\eta^{(j)}_{k_j}&\frac{i^{k_N}\sqrt{N}}{{2}}\Omega J_{k_N}(\frac{2\mu_1}{\omega})\\
0&0&0&\ldots&- \frac{i^{-k_N}\sqrt{N}}{{2}}\Omega J_{k_N}(\frac{2\mu_1}{\omega})&\sum_{j=1}^N\eta^{(j)}_{k_j}
\end{array}\right).
\end{equation}

 The time-dependent solutions corresponding to eigensolutions of Eq.~(\ref{eq:eigen}) with eigenvalue~$\omega_{\ell}$ read:
\begin{equation}
\label{eq:loesung}
\left(
\begin{array}{c}
{a}_0(t)\\
{a}_1(t)\\
{a}_2(t)\\
\ldots\\
{a}_N(t)
\end{array}\right)
=
\left(
\begin{array}{c}
\widetilde{a}^{(\ell)}_0\exp[-i\omega_{\ell}t]\\
\widetilde{a}^{(\ell)}_1\exp\left[-i\left(\omega_{\ell}-\eta_{k_1}^{(1)}\right)t\right]\\
\widetilde{a}^{(\ell)}_2\exp\left[-i\left(\omega_{\ell}-\eta_{k_1}^{(1)}-\eta_{k_2}^{(2)}\right)t\right]\\
\ldots\\
\widetilde{a}^{(\ell)}_N\exp\left[-i\left(\omega_{\ell}-\sum_{j=1}^N\eta_{k_j}^{(j)}\right)t\right]
\end{array}\right)
\end{equation}
\end{widetext}

\section{\label{sec:analytic}Analytic calculations}
This section demonstrates that it is possible to obtain analytic results for the particle transfer for fractional photon-assisted tunneling. Starting with the matrix~(\ref{eq:matrix4}) for the experimentally relevant~\cite{CheinetEtAl08} case of $N=2$ particles, one then obtains a set of eigenvalues:
\begin{equation}
\omega_1=0\quad\omega_{2/3}=\frac{\sigma_{k'}}{2}\pm\sqrt{\frac{\sigma_{k'}^2}{4}+\Omega_1^2+\Omega_2^2}
\end{equation}
and their corresponding eigenvectors:
\begin{equation}
\label{eq:vectors}
\widetilde{v}^{(1)}=\left(\begin{array}{c}
-i\frac{\Omega_2}{\Omega_1}\\
0\\
1 
\end{array}\right),\;
\widetilde{v}^{(2)}=\left(\begin{array}{c}
i\frac{\Omega_1}{\Omega_2}\\
-i\frac{\omega_{2}}{\Omega_2}\\
1 
\end{array}\right),\;
\widetilde{v}^{(3)}=\left(\begin{array}{c}
i\frac{\Omega_1}{\Omega_2}\\
-i\frac{\omega_{3}}{\Omega_2}\\
1 
\end{array}\right).
\end{equation}

These results are now inserted into the ansatz~(\ref{eq:ansatz}) and thus yield three linearly independent solutions ${v}^{(1)}(t)$,  ${v}^{(2)}(t)$ and  ${v}^{(3)}(t)$. Any state of the system $\left|\Psi(t)\right\rangle$ can be expressed as a linear combination $\left|\Psi(t)\right\rangle=b_1v^{(1)}(t)+b_2v^{(2)}(t)+b_3v^{(3)}(t)$. The complex coefficients $b_j$ are obtained from the initial condition $\left|\Psi(0)\right\rangle\equiv\left|0\right\rangle$. The time dependent transfer~(\ref{eq:transfer}), can then be expressed in terms of the amplitudes $a_j(t)$ which depend on the time-dependent solutions corresponding to the eigenvectors~(\ref{eq:vectors}):
\begin{eqnarray}
\label{eq:imbalance2}
P_{\rm trans}&=&\frac{2\left|a_2(t)\right|^2+\left|a_1(t)\right|^2}{2}\nonumber\\
&=&\frac12\left(2\left|\sum_{j=1}^3b_j{v}^{(j)}_3(t)\right|^2\right.\nonumber\\
&&+\left.\left|\sum_{j=1}^3b_j{v}^{(j)}_2(t)\right|^2\right)
\end{eqnarray}
In order to achieve perfect transfer one finds $\sigma_{k'}=0$ implying $\omega_{2/3}=\pm\sqrt{\Omega_1^2+\Omega_2^2}$. Now in the expression for the transfer all phase factors only contain frequencies, which are integer multiples of $\omega_{2/3}$, thus leading to a retrieval of the initial state for $\omega_2t=2\pi$. Best transfer is therefore expected to occur for 
\begin{equation}
\label{eq:tstar}
t^*=\frac{\pi}{\sqrt{\Omega_1^2+\Omega_2^2}}.
\end{equation}
From the definition~(\ref{eq:sigk}) one finds $\sigma_{k'}=0$ for $k'=0$ and $\kappa=\omega/4$.

To account for experimentally realistic conditions, that is adjusting the interaction strength $\kappa/\Omega$ via modulations of the depth of the wells, we investigate how sensitive perfect transfer is to changes in the interaction parameter $\kappa$. It is reasonable to consider changes within the range $0\leq\kappa\leq\omega/2$ as $k'$ will remain zero. Inserting these values into the particle transfer~(\ref{eq:imbalance2}) leads to an analytic expression for the transfer:
\begin{eqnarray}
\label{eq:imbalance3}
P_{\rm trans}&=&\frac{1}{2}\left[\left|P(0,t)-P(\omega_2,t)-P(\omega_3,t)\right|^2\right.\\
&-&\left.\left|-i\frac{\Omega_2}{\Omega_1}P(0,t)-i\frac{\Omega_1}{\Omega_2}\left(P(\omega_2,t)+P(\omega_3,t)\right)\right|^2\right],\nonumber
\end{eqnarray}
where
\begin{equation}
P(\omega,t)\equiv\frac{i\exp(-i\omega t)}{\sqrt{2+\frac{\Omega_1^2}{\Omega_2^2}+\frac{\Omega_2^2}{\Omega_1^2}+2\frac{\omega^2}{\Omega_1^2}+2\frac{\omega^2}{\Omega_2^2}+\frac{\omega^4}{\Omega_1^2\Omega_2^2}}}\;.
\end{equation}
Equation~(\ref{eq:imbalance3}) is displayed in Fig.~\ref{fig:transfer}.
\end{appendix}

%

\end{document}